# Monkeys get a silver in Abstract Art Olympics

Mikhail Simkin

This past summer I discussed [1] the recent comparative study [2] of abstract art and animal art. Researches had shown to art students pairs of paintings: one by a renowned abstract expressionist and another by a child, a monkey, an ape, or an elephant. Researches asked the students which painting is better. In 67.5% of the cases art students answered that the painting by an artist is better. Because the figure is above 50%, researches concluded that their findings "challenge the common claim that abstract expressionist art is indistinguishable from (and no better than) art made by children." I recalled [1] another experiment, where the subjects compared weights. In 72% of the trials, they perceived a 100-gram weight to be heavier than 96-gram weight, a higher success rate than in the art experiment. I concluded that the difference in artistic weight of abstract artists and animals is less than 4%. I also compared [1] the results of the art experiment with the outcomes of the games of chess-players of different strength and concluded that abstract artists are in the same player category with children and animals. My commentary got attention from the media [3], often critical, and I responded to criticism [4]. However, some of the objections have merit. To compare paintings one should consider many parameters, while to compare weights – only one. More parameters give more reasons for disagreement between judges. In chess, there no judges at all, since the rules of the game determine its outcome. However, there is a sport where judges decide the results of the competitions: Figure Skating. Now let us look at statistics.

In Olympics Figure skating program the skaters compete for four sets of medals in four categories: men, ladies, pairs, and ice dancing. Competition consists of several segments. In men, ladies and pairs categories there are only two segments: short program and free skating. In ice dancing category there are four segments. The table shows the results [5] of the 2002 Olympics in Salt Lake City. Under the old 6.0 system [6], nine judges independently judge each segment. Every judge gives every skater a score, which depends on many parameters of performance (including aesthetic parameters, just like in art). Next, they look at all the scores given by a particular judge and rank the skaters from the highest score to the lowest. The skater, who got the highest number of first places, wins the segment. Out of the remaining skaters, the one who got the highest number of either first or second places gets the second place and so on. For example, in Ladies Free Skating segment (see the Table), Hughes got five first places, more than anyone else, and won the segment. Slutskaya got four first and two second places that is six first/second places and got the second place in the segment. To determine the final placement, they sum the places the skaters got in all segments multiplied by corresponding weight-factors. Free Skate's weight is 2/3 and the Short Program's – 1/3. The skater with the lowest total is the winner. If two skaters are tied, the winner is who places higher in the Free Skate. Hughes got the first place in Free Skate and the fourth place in Short Program. Her total is 2/3 + 4/3 = 2. Slutskaya got second places in both segments and her total is 2, same as Hughes's. Since Hughes placed first in Free Skate, she won the Gold and Slutskaya got the Silver.

Results depend on the arbitrary prescription for breaking ties and arbitrary weights of different segments. If the weights were not 2/3 and 1/3, but 3/5 and 2/5, Slutskaya, not Hughes, would win the Gold. The aim of this commentary is not to criticize the 6.0 scoring system (they do not even use it today: they replaced it with a far more muddled one). The aim is to show that that the system is robust. When we changed weights, former silver medalist became gold medalist, not someone from the end of the list. One cannot propose any reasonable modification of the scoring system, which will propel the tenth skater to the first place. The same is true when instead of changing some of the rules we remove the rankings given by some of the judges.

**Table 1.** Figure Skating results in 2002 Olympics in Salt Lake City. The rightmost columns show the numbers and percentages of judges who placed Gold medalist over Silver medalist, Gold medalist over Bronze medalist and Silver medalist over Bronze medalist**.** I treated the Pairs category as if Sale and Pelletier got a Silver medal, but discarded the placements given by the disqualified judge.

| Category / Medalists | Segment | Medalist skaters | Place in the segment | Places given by each of 9 judges | Gold over Silver | | Gold over Bronze | | Silver over Bronze | |
|---|---|---|---|---|---|---|---|---|---|---|
| **Men** <br> G Yagudin <br> S Plushenko <br> B Goebel | Short Program | Yagudin <br> Goebel <br> Plushenko | 1 <br> 3 <br> 4 | 1 1 1 1 1 1 1 1 1 <br> 2 5 2 3 4 2 2 3 4 <br> 5 3 5 5 3 4 5 5 3 | 9 | 100% | 9 | 100% | 3 | 33% |
| | Free Skating | Yagudin <br> Plushenko <br> Goebel | 1 <br> 2 <br> 3 | 1 1 1 1 1 1 1 1 1 <br> 2 2 2 2 2 2 2 2 2 <br> 5 4 3 3 3 3 3 3 3 | 9 | 100% | 9 | 100% | 9 | 100% |
| **Ladies** <br> G Hughes <br> S Slutskaya <br> B Kwan | Short Program | Kwan <br> Slutskaya <br> Hughes | 1 <br> 2 <br> 4 | 1 2 1 1 1 2 2 2 1 <br> 3 1 2 2 3 1 1 1 3 <br> 6 10 4 5 5 5 5 4 4 | 0 | 0% | 0 | 0% | 4 | 44% |
| | Free Skating | Hughes <br> Slutskaya <br> Kwan | 1 <br> 2 <br> 3 | 1 4 3 4 1 2 1 1 1 <br> 3 1 1 1 4 1 2 3 2 <br> 2 3 2 2 2 3 3 2 3 | 5 | 56% | 6 | 67% | 6 | 67% |
| **Pairs** <br> G Berezhnaya / Sikharulidze <br> G Sale / Pelletier <br> B Shen / Zhao | Short Program | Berezhnaya / Sikharulidze <br> Sale / Pelletier <br> Shen / Zhao | 1 <br> 2 <br> 3 | 1 1 1 1 1 2 1 2 1 <br> 2 2 2 2 2 1 2 1 2 <br> 3 3 3 3 3 3 3 3 3 | 7 | 78% | 9 | 100% | 9 | 100% |
| | Free Skating | Berezhnaya / Sikharulidze <br> Sale / Pelletier <br> Shen / Zhao | 1 <br> 1 <br> 3 | 1 1 2   1 2 1 2 2 <br> 2 2 1   2 1 2 1 1 <br> 3 3 3   3 3 3 3 3 | 4 | 50% | 8 | 100% | 8 | 100% |
| **Ice Dancing** <br> G Anissina / Peizerat <br> S Lobacheva / Averbukh <br> B Fusar Poli / Margaglio | 1st Compulsory Dance | Anissina / Peizerat <br> Lobacheva / Averbukh <br> Fusar Poli / Margaglio | 1 <br> 2 <br> 3 | 1 1 1 2 1 1 1 1 1 <br> 2 2 2 3 2 2 2 3 4 <br> 4 4 3 1 3 3 3 2 2 | 9 | 100% | 8 | 89% | 6 | 67% |
| | 2nd Compulsory Dance | Anissina / Peizerat <br> Lobacheva / Averbukh <br> Fusar Poli / Margaglio | 1 <br> 2 <br> 3 | 1 1 2 1 1 1 1 1 1 <br> 2 3 4 3 2 2 3 2 3 <br> 3 2 1 2 3 3 2 3 4 | 9 | 100% | 8 | 89% | 5 | 56% |
| | Original Dance | Anissina / Peizerat <br> Lobacheva / Averbukh <br> Fusar Poli / Margaglio | 1 <br> 2 <br> 3 | 2 1 1 1 1 1 1 1 1 <br> 1 3 3 2 2 2 2 2 2 <br> 3 2 2 3 4 3 3 4 4 | 8 | 89% | 9 | 100% | 7 | 78% |
| | Free Dance | Anissina / Peizerat <br> Lobacheva / Averbukh <br> Fusar Poli / Margaglio | 1 <br> 2 <br> 3 | 2 2 1 2 2 1 1 1 1 <br> 1 1 2 1 1 2 2 2 2 <br> 3 3 4 4 3 3 4 4 3 | 5 | 56% | 9 | 100% | 9 | 100% |
| | | | average | | 6.5 | 73% | 7.5 | 84% | 6.6 | 74% |

If you look at the Pairs Free Skate results, you notice that the placements given by the fourth judge are missing. This is because of the 2002 Olympic figure skating scandal [7]. Initially Sale and Pelletier received the second place in Free Skate and a Silver medal. However, North American news media got angry at this decision. For example, Christine Brennan in her USA Today article "No defense for bad judgment" wrote that she cannot defend the sport of Figure Skating after "Jamie Sale and David Pelletier skated one of the great performances in Olympic history, only to have the gold medal taken from them by five misguided people on the nine-person judging panel." International Skating Union declared one judge corrupt and discarded her scores. Sale and Pelletier became tied with Berezhnaya and Sikharulidze in Free Skate segment. Based on that, the officials awarded a second set of Gold medals in Pairs category. One can question this decision because Berezhnaya and Sikharulidze still had won the short program and therefore their total is 1, while Sale and Pelletier's total is 4/3. To properly award Gold to Sale and Pelletier, one would have to disqualify one more misguided judge. Note by the way, that even if we disqualify all five misguided judges, Berezhnaya and Sikharulidze would fall down only to the second place, not to the twentieth. So much of a scandal.

The aim of the preceding passage is not to ridicule the fight against corruption. On contrary, the aim is to encourage American news media to uncover corruption in more places. For each segment, I computed the percentage of the judges who placed Gold medalist over Silver medalist (see the Table). The average over all 10 segments is 73%. This exceeds the 68% result I started the article with. Thus, the difference between grand masters of abstract art and apes is less than the difference between Gold and Silver medalists. The Figure Skating scandal was about who of world-class skaters should get Gold and who should get Silver. Though sport critics questioned Olympics Gold medals of Berezhnaya and Sikharulidze, nobody questioned the two Gold medals they got in two World Championships. In the abstract art experiment, the paintings of renowned artists did not compete with the art of world champions. They competed with paintings any child or monkey can do. Why are the art critics silent? Why nobody asks how corrupt the judges that decide which of the paintings will cost millions of dollars must be?